\UseRawInputEncoding
\documentclass[twocolumn]{aastex631}

\newcommand\hst{{\it HST}}
\newcommand\jwst{{\it JWST}}

\begin{document}

\title{A First Look into the Nature of JWST/MIRI $7.7\mu$m Sources from SMACS 0723}

\correspondingauthor{Edoardo Iani}
\email{iani@astro.rug.nl, E.Iani@rug.nl}

\author[0000-0001-8386-3546]{Edoardo Iani}
\affiliation{Kapteyn Astronomical Institute, University of Groningen, P.O. Box 800, 9700AV Groningen, The Netherlands}

\author[0000-0001-8183-1460]{Karina I. Caputi}
\affiliation{Kapteyn Astronomical Institute, University of Groningen, P.O. Box 800, 9700AV Groningen, The Netherlands}
\affiliation{Cosmic Dawn Center (DAWN), Copenhagen, Denmark}

\author[0000-0002-5104-8245]{Pierluigi Rinaldi}
\affiliation{Kapteyn Astronomical Institute, University of Groningen, P.O. Box 800, 9700AV Groningen, The Netherlands}

\author[0000-0002-5588-9156]{Vasily Kokorev}
\affiliation{Kapteyn Astronomical Institute, University of Groningen, P.O. Box 800, 9700AV Groningen, The Netherlands}
\affiliation{Niels Bohr Institute, University of Copenhagen, Blegdamsvej 17, DK2100 Copenhagen Ø, Denmark}
\affiliation{Cosmic Dawn Center (DAWN), Copenhagen, Denmark}



\begin{abstract}

Until now, our knowledge of the extragalactic Universe at mid-IR wavelengths ($> 5 \, \rm \mu m$) was limited to rare active galactic nuclei (AGN) and the brightest normal galaxies up to $z\sim3$. The advent of the \jwst  ~with its Mid-Infrared Instrument (MIRI) will revolutionise the ability of the mid-IR regime as a key wavelength domain to probe the high-$z$ Universe. In this work we present a first study of \jwst~MIRI $7.7 \, \rm \mu m$ sources selected with $> 3 \, \sigma$ significance from the lensing cluster field SMACS~J0723.3-7327. We model their spectral energy distribution fitting with 13 \jwst ~and \hst ~broad bands, in order to obtain photometric redshifts and derived physical parameters for all these sources. We find that this  $7.7 \, \rm \mu m$ galaxy sample is mainly composed of normal galaxies up to $z=4$ and has a tail of about 2\% of sources at higher redshifts to $z\approx 9-10$.
The vast majority of our galaxies have $[3.6]-[7.7]<0$ colours and very few of them need high dust extinction values ($A_V=3 -6$ mag) for their SED fitting. The resulting lensing-corrected stellar masses span the range $10^7-10^{11} \, \rm M_\odot$. Overall, our results clearly show that the first MIRI $7.7 \, \rm \mu m$ observations of deep fields are already useful to probe the high-redshift Universe and suggest that the deeper $7.7 \, \rm \mu m$ observations to be available very soon will open up, for the first time, the epoch of reionisation at mid-IR wavelengths.

\end{abstract}



\section{Introduction} \label{sec:intro}

Our knowledge of the extragalactic background sky has significantly improved over the past decades with observations from space and ground-based observatories operating at different wavelengths. In the mid-infrared (mid-IR) regime ($3-30 \, \rm \mu m$), in particular, a major leap forward has been achieved with the {\it Spitzer Space Telescope} \cite[{\it Spitzer},][]{Werner+04}, which has shown the important contribution of mid-IR wavelengths to the general extragalactic background light \cite[][]{Dole+06}.

Moreover, {\it Spitzer} observations allowed us for the first time to reveal the high-redshift Universe at mid-IR wavelengths \cite[see e.g.][and references therein]{Bradac+20}.  Using the Infrared Array Camera \cite[IRAC, ][]{Fazio+04} 3.6 and 4.5~$\rm \mu m$, numerous works have detected and studied thousands of high-$z$ galaxies up to $z\sim6$ \cite[e.g.,][]{Caputi+15, Caputi+17, Davidzon+17} and even hundreds of galaxies at higher redshifts \cite[e.g.,][]{Stefanon+21}. 

At wavelengths $\lambda> 5 \, \rm \mu m$, the decreasing instrumental performance coupled with the increasing background noise and the limited lifetime of the cryogenic systems implies that reaching high sensitivities is much more difficult. Indeed, IRAC could reach at most 21-22 mag ($5 \, \sigma$) in its extragalactic surveys at 5.8 and 8.0~$\rm \mu m$ \cite[][]{Euclid+22}. 
This resulted in a very limited utility of these photometric bands for the study of galaxies at $z>2$, except for rare, mid-IR bright active galactic nuclei \cite[AGN, e.g.][]{Stern+05, Eisenhardt+12}.
Similarly, at even longer wavelengths, observations with The Multiband Imaging Photometer for {\it Spitzer} \cite[MIPS, ][]{Rieke+04} were able to find sources up to $z \sim 3$ \cite[e.g.,][]{LeFloch+09,Huynh+10}, but such high redshifts were very rare and above $z\sim3$ all the detected sources virtually hosted AGNs.

The advent of the \jwst\ with its Mid-Infrared Instrument \cite[MIRI;][]{Rieke+15, Wright+15} is expected to radically change the ability of mid-IR astronomy to study the high-$z$ Universe. The goal of this work is to showcase the major leap forward produced by MIRI from its first observations of extragalactic fields. 

In this paper, we adopt a flat $\Lambda$-CDM cosmology with $\Omega_\Lambda = 0.7$, $\Omega_m = 0.3$, and $H_0 = 70\ \text{km/s/Mpc}$.
All magnitudes are in the AB photometric system \cite[][]{Oke74}.

\section{Target and Data} \label{sec:data}
Our target is SMACS J0723.3-7327 (hereafter
SMACS 0723), a galaxy lensing cluster located in the southern celestial hemisphere (${\rm R.A.}(J2000.0) = 07^h 23^m 19.5^s$, ${\rm Dec.}(J2000.0) = -73^\circ 27' 15.6''$) at $z = 0.39$ \citep{Martinez-Aviles+18}.
Here we describe the datasets used in our study, which comprise observations from \jwst\ (MIRI, NIRCam) and the {\it Hubble Space Telescope} (\hst).
We also briefly describe the lensing model we adopted to retrieve the physical parameters of our background sources.

\subsection{\jwst\ imaging} \label{subsec:jwst}
The \jwst\ observations of SMACS 0723 (ID proposal: 2736\footnote{\url{https://www.stsci.edu/jwst/phase2-public/2736.pdf}}, PI: K. M. Pontoppidan) were publicly released as part of the \jwst\ Early Release Observations (EROs) program on the 13th of July 2022.
The observations were carried out with the \jwst\ Near Infrared Camera \cite[NIRCam,][]{Rieke+05}, the Mid-Infrared Instrument \cite[MIRI,][]{Rieke+15, Wright+15}, the Near Infrared Imager and Slitless Spectrograph \cite[NIRISS,][]{Willott+22} and the Near Infrared Spectrograph \cite[NIRSpec,][]{Jakobsen+22}.
In the following, we briefly describe the MIRI and NIRCam observations we exploited in our work.

\subsubsection{MIRI} \label{subsec:miri}
The MIRI instrument on-board of \jwst\ observed the central region of SMACS 0723 with the wide-band filters F770W, F1000W, F1500W and F1800W.
The observations were carried out on the 14th of June 2022, and covered an area of $112.6''\times 73.5''$, i.e. the MIRI field-of-view (FoV). In this work, we only make use of the F770W image.

We downloaded the fully-reduced MIRI images available at the Mikulski Archive for Space Telescopes (MAST)\footnote{\url{https://mast.stsci.edu/portal/Mashup/Clients/Mast/Portal.html}}.
From a visual inspection we found strong background patterns such as vertical stripes and gradients. 
The intensity of these features varies significantly among filters, thus suggesting that the calibration files available for the MIRI data reduction are not performing well.
To improve the background cleaning and homogenisation, we decided to re-run the \jwst\ pipeline (version 1.6.1\footnote{CRDS version 11.16.5, CRDS context \texttt{jwst\_0916.pmap}.}) implementing an ad-hoc algorithm able to remove systematics from the fully calibrated exposures generated by the pipeline at the end of Stage 2 and before the final mosaicing performed in Stage 3.
The introduction of this extra-step during the data reduction does not only impact on the images cosmetic. 
In fact, by running and comparing \texttt{SExtractor} catalogues obtained with the same configuration file and based on both the final image obtained with and without the extra-step, we found a substantial reduction of spurious detections (especially along the vertical stripes).
Contextually, the extra-step helped in maximising the number of real sources detected. 
Ultimately, to ensure that the extra-cleaning of the background did not impact on our magnitude estimates, we compared the magnitudes of bright ($< 24~ {\rm mag}$) sources.
We found no systematic offset between the two estimates. 

A comparison between the MIRI and \hst\ images highlighted an astrometric offset between the two dataset.
We decided to register the MIRI observations to the \hst\ astronomical coordinate system.
To do so, we generated catalogues of sources for all MIRI images and a stacked image of the \hst\ observations via the Source Extractor software \cite[\texttt{SExtractor},][]{Bertin+96}.
We then matched the catalogues, minimising the offsets between the detected sources.
We found a median offset of $\Delta{\rm R.A.} = 0.50\pm0.27$ arcsec and $\Delta {\rm Dec.} = 1.28\pm0.08$ arcsec.

Finally, we resampled the MIRI images to \hst\ via the \texttt{Python} library \texttt{reproject}\footnote{\url{https://reproject.readthedocs.io/en/stable/}}, an affiliated package of \texttt{Astropy} \cite[][]{astropy13, astropy18}.

\subsubsection{NIRCam} \label{subsec:nircam}
The NIRCam instrument on-board of \jwst\ observed SMACS 0723 with the wide-band filters F090W, F150W, F200W, F277W, F356W and F444W.
The observations were carried out on the 6th of June 2022.
NIRCam is constituted by two modules (A and B) pointing to adjacent fields of view and covering an area of $2.2'\times 2.2'$ (each).
For the observations of SMACS 0723, only module B targeted the centre of the cluster, while module A provided observations for a parallel field.

We downloaded the fully-reduced NIRCam images available at MAST.
Differently from MIRI, a visual inspection of the dataset revealed a good quality of the automatic reduction of NIRCam images by the \jwst\ pipeline.
Nonetheless, following the release of the new photometric zero-points of NIRCam (CRDS context \texttt{jwst\_0989.pmap}, 3rd of October 2022), we corrected the dataset at our disposal accordingly, see Table~\ref{tab:table_photometry}.

Also for the NIRCam observations, we used \texttt{SExtractor} to correct the astrometry of the images and register them to the \hst\ dataset.
In this case, we found a median offset of $\Delta{\rm R.A.} = 1.03\pm0.05$ arcsec and $\Delta{\rm Dec.} = 0.01\pm0.02$ arcsec.
We finally resampled the NIRCam images to the \hst\ spatial sampling.

\subsection{\hst\ imaging} \label{subsec:hst}
SMACS 0723 was observed with \hst\ as part of The Reionization Lensing Cluster Survey \cite[RELICS,][]{Coe+19} that targeted 41
massive galaxy clusters  at redshift $z \sim 0.2 - 1.0$, including 21 of the 34 most massive known according to {\it Planck} \cite[][]{Planck+11}.
The observations of SMACS 0723 were carried out with the ACS/WFC F435W, F606W and F814W filters, and with the WFC3/IR F105W, F125W, F140W and F160W filters. 

We downloaded the fully-reduced \hst\ images from the RELICS repository\footnote{\url{https://archive.stsci.edu/missions/hlsp/relics/SMACS 0723-73/}}. 
Although we used as reference the RELICS astrometry in our study, we found an offset of $\Delta\rm R.A. = -0.53\pm0.03$ arcsec and $\Delta\rm Dec. = 0.32\pm0.01$ arcsec when comparing it with the {\it Gaia} DR3 catalogue \cite[][]{gaia_dr3}.

\subsection{Lensing model} \label{subsec:lensing}
To derive the de-magnified stellar mass of our targets (see Section~\ref{subsec:stellar_mass}), we adopted the recent lensing model of SMACS 0723 by \cite{Caminha+22}.
The lensing model was obtain using the software \texttt{lenstool} \cite[][]{Kneib+96, Jullo+07, Jullo+09} to model the mass distribution of the galaxy cluster. 
In particular, the fiducial mass model was constructed by means of an elliptical cluster-scale dark matter halo, a truncated spherical isothermal mass profile for each cluster galaxy member, and an external shear. 
The positions of multiple images detected in the \jwst/NIRCam images and (when available) spectroscopic redshifts from the Multi-Unit Spectroscopic Explorer \cite[MUSE,][]{Bacon+10} were adopted to constrain the model. 
We refer the reader to \cite{Caminha+22} for a more detailed description of the SMACS 0723 lensing model.

\section{Photometric catalogue and sample selection} \label{sec:sample_catalogue}

After registering and resampling all the images to a same system of coordinates, we constructed our sample. To do so, we extracted photometric catalogues by making use of the software \texttt{SExtractor} \cite[][]{Bertin+96}. We ran \texttt{SExtractor} in dual-mode, using for detection the MIRI  $7.7 \mu {\rm m}$ image and measuring the photometry in each of the 14 filters considered here (Table~\ref{tab:table_photometry}).    To identify the $7.7\, \rm \mu m$ sources, we adopted a {\it hot-mode} configuration \cite[][]{Galametz+13} and used the weights extension of the MIRI images to improve the rejection of spurious sources.

We measured each source photometry adopting circular apertures of 1 arcsec diameter via the \texttt{MAG\_APER} task.
We corrected these flux estimates for aperture effects $f_{\rm aper}$, see Table~\ref{tab:table_photometry}. 
For NIRCam and MIRI, we estimated the aperture correction factor from the \texttt{Python} software \texttt{WebbPSF} \cite[][]{Perrin+14}.

Whenever dealing with bright and extended sources, circular aperture photometry fails in estimating the galaxy total magnitude. 
Similarly, it is not suitable for retrieving the photometry of sources significantly stretched by the cluster lensing effect. 
For these reasons, for all the detected sources in our catalogue, we also measured Kron aperture photometry \cite[][]{Kron80} via the \texttt{SExtractor} task \texttt{MAG\_AUTO}.
In our final catalogue we adopt the aperture-corrected \texttt{MAG\_APER} in all cases with $>23$~mag, while at $<23$~mag we kept the brightest measurement between the aperture-corrected \texttt{MAG\_APER} and \texttt{MAG\_AUTO}.   

Finally, we corrected our photometry for Galactic extinction.
This was done by means of the \texttt{Python} packages \texttt{extinction} and \texttt{dustmaps} \cite[][]{Green+18}, and assuming the Galactic extinction curve by \cite{Cardelli+89} with $R_V = 3.1$.
We derived values for the colour excess E(B-V) in agreement with\footnote{\url{ https://irsa.ipac.caltech.edu/applications/DUST/}} those by \cite{Schlafly+11}.

Since \texttt{SExtractor} has been found to generally underestimate photometric errors \cite[e.g.][]{Sonnett+13}, we manually set a minimum error of 0.05~mag in all the available filters. 
This is the value commonly considered as minimum systematic error for \hst\ imaging data and we adopted it for the \jwst\ observations too.

Finally, we computed flux upper-limits in all cases of a  \texttt{SExtractor} non-detection. 
To do so, we masked all sources based on the \texttt{SExtractor} segmentation map and estimated the local background flux by randomly placing 1 arcsec diameter apertures in the background and at a maximum distance of 10 arcsec of the non-detected source. After applying a $3\sigma$ clipping (until convergence) on the distribution of fluxes retrieved, we obtained the standard deviation ($1\sigma$) of the local background. 
In our final catalogue, we consider the $3\sigma$ background level as upper-limit on the source flux. 

To assess the quality of our photometry, we compared it with the RELICS catalogue\footnote{\url{https://archive.stsci.edu/missions/hlsp/relics/SMACS 0723-73/catalogs/}}. 
The agreement with the RELICS magnitudes allowed us to validate our own photometric measurements.
We find a difference between our photometry and the RELICS catalogue $\leq 0.1$ mag (median) for all the {\it HST} filters but for ACS/F435W ($\simeq 0.6$ mag). This wider offset reflects the shallower depth of the filter ($\simeq 23.6$ mag at 5$\sigma$) and the fact that our selected sample has F435W magnitudes only between $22-24$. 
In addition, we compared our photometry with that provided in the SMACS 0723 photometric catalogue obtained by \cite{Kokorev+22}. As this catalogue includes IRAC photometry at 3.6 and 4.5~$\rm \mu m$, we could directly asses the quality of the NIRCam absolute flux calibration at its longest wavelengths. The absence of {\it Spitzer}/IRAC observations in channel 4 ($8.0~\mu{\rm m}$) prevented us from verifying the quality of the absolute flux calibration for the MIRI F770W filter.

To construct a reliable sample of sources, we limit our analysis to objects having at least a $3\sigma$ detection both at $7.7\mu{\rm m}$ (MIRI F770W filter) and $3.6\mu{\rm m}$ (NIRCam F356W filter).  We have excluded 11 Galactic stars identified from the {\it Gaia} DR3 catalogue \cite[][]{gaia_dr3} and other 4 close sources whose reliability and/or photometry are compromised by the bright stellar light.

\begin{figure}
    \centering
    \includegraphics[width=\columnwidth]{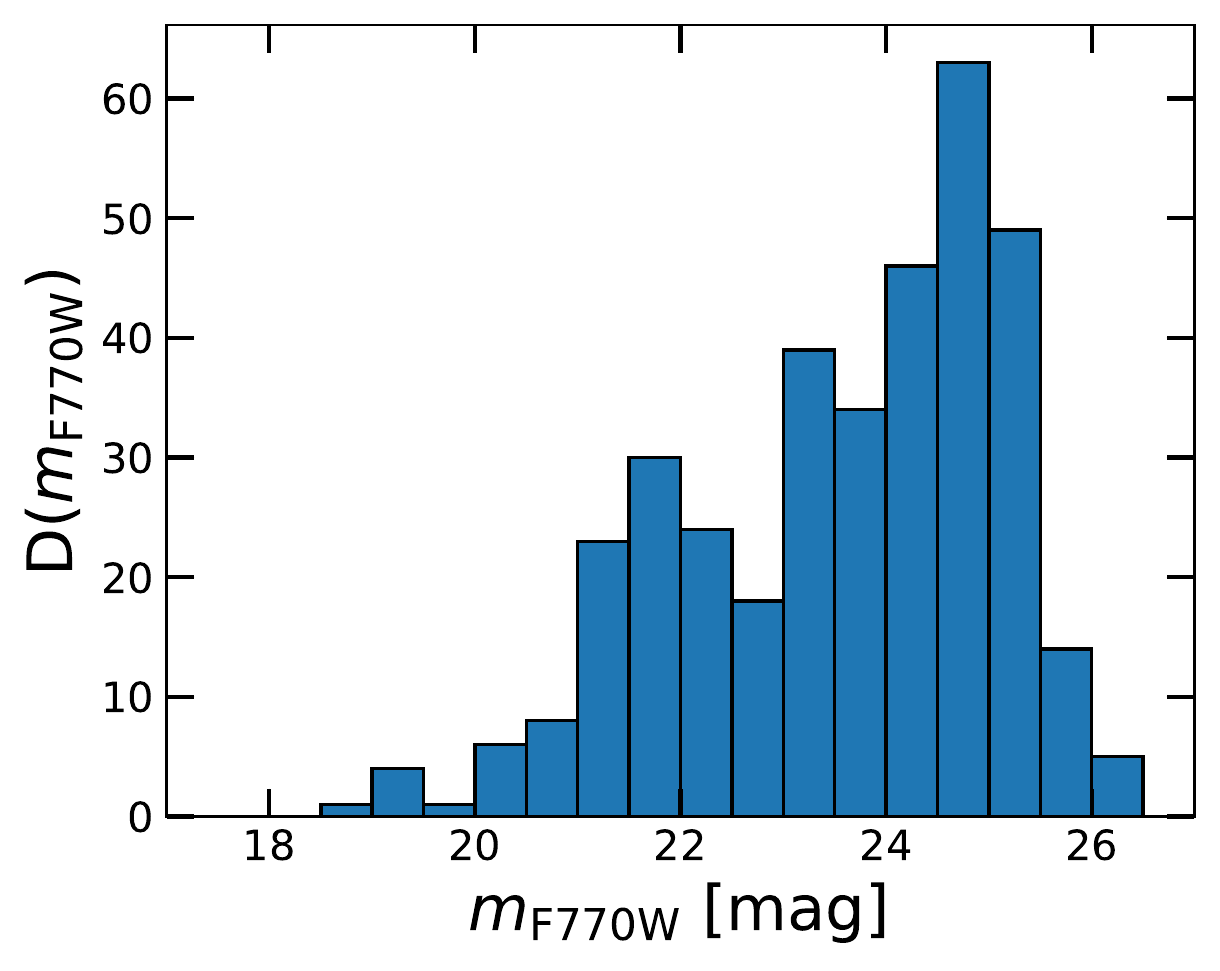}
    \caption{Distribution of \jwst ~F770W magnitudes for our sample of 7.7~$\rm \mu m$-selected galaxies with $S/N \rm (F770W)>3$ and $S/N \rm (F356W)>3$ in SMACS 0723. There are 361 sources in total.}
    \label{fig:mag_dist}
\end{figure}

\begin{figure*}
    \centering
    \includegraphics[width=\textwidth]{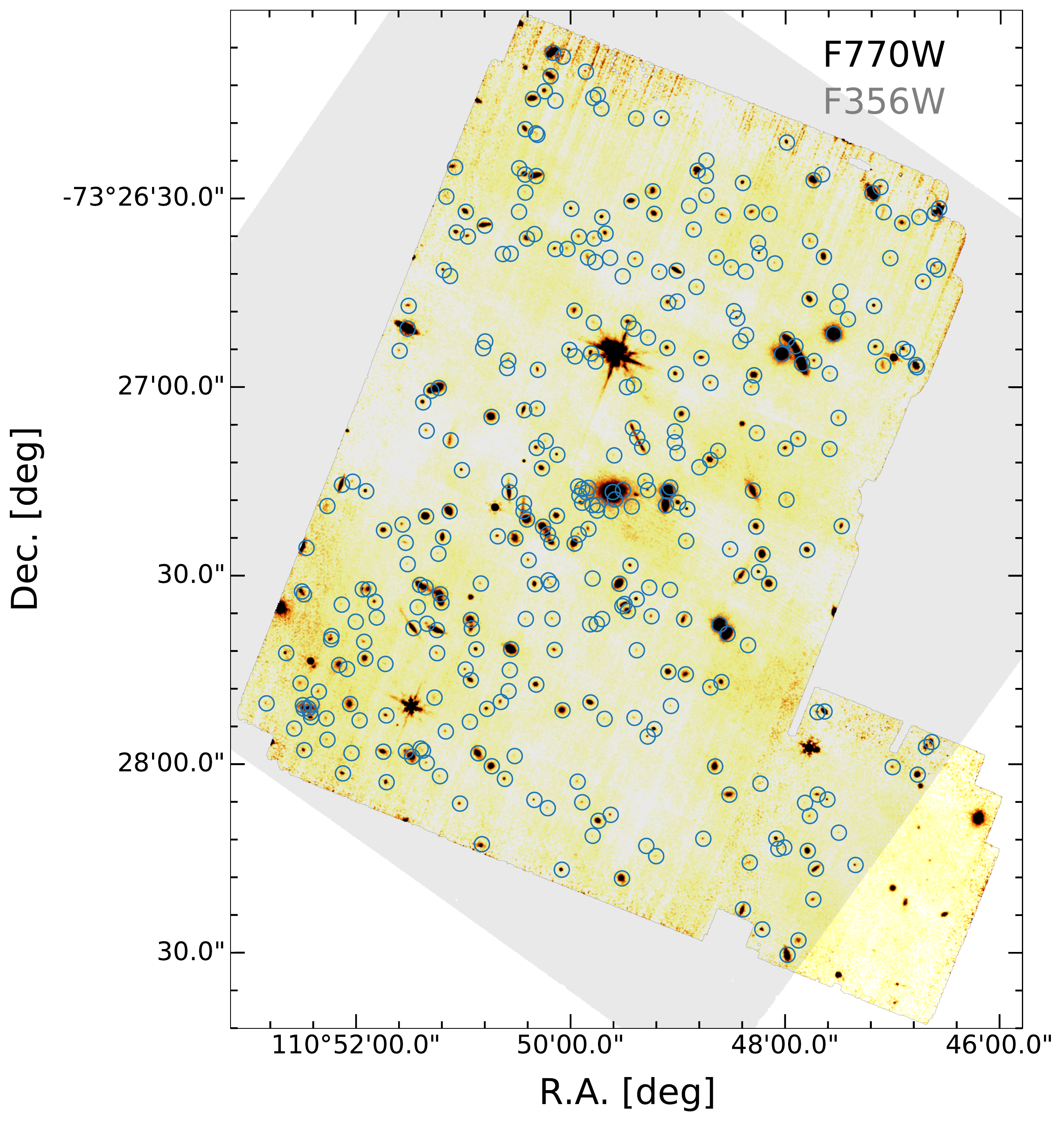}
    \caption{Layout of the \jwst/MIRI $7.7\mu{\rm m}$ (F770W) image for SMACS 0723. The open blue circles indicate the spatial position of our final sample of 341 sources having a 3$\sigma$ detection in both F770W and the \jwst/NIRCam F356W filter. The grey-shaded area shows the region of SMACS 0723 covered by the \jwst/NIRCam B module.}
    \label{fig:map_srcs}
\end{figure*}

As a result, we end up with a final sample constituted by 361 sources.
We present the F770W magnitude distribution of our sample in Figure~\ref{fig:mag_dist} and show their spatial distribution in the SMACS 0723 F770W image in Figure~\ref{fig:map_srcs}. The F770W magnitude distribution shows that the number counts increase up to $\sim 25$~mag and then we quickly lose completeness at higher magnitudes.  This is about 2.5 magnitudes deeper than any IRAC 8~$\mu m$ image available for extragalactic fields.

\begin{deluxetable}{lcccccr}
\tablenum{1}
\tablecaption{HST and JWST data for SMACS 0723 analysed in this work}
\tablewidth{0pt}
\tablehead{
\colhead{Instrument} & \colhead{Filter} & \colhead{${\rm ZPT}_{\rm AB}$} & \colhead{$f_{\rm ext}$} & \colhead{$f_{\rm aper}$} & \colhead{$f_{\rm ZPT}$} & \colhead{$t_{\rm exp}$}\\
 & & [mag] & & & & [ks]}
\startdata
\hst/ACS     & F435W  & 25.664 & 2.047 & 1.103 & - & 2.2 \\ 
\hst/ACS     & F606W  & 26.500 & 1.649 & 1.092 & - & 2.3  \\
\hst/ACS     & F814W  & 25.945 & 1.377 & 1.094 & - & 2.5  \\
\jwst/NIRCam & F090W  & 26.705 & 1.291 & 1.082 & 1.051 & 30.1 \\
\hst/WFC3 IR & F105W  & 26.269 & 1.223 & 1.136 & - & 1.5  \\
\hst/WFC3 IR & F125W  & 26.230 & 1.165 & 1.157 & - & 0.8   \\        
\hst/WFC3 IR & F140W  & 26.452 & 1.138 & 1.166 & - & 0.8   \\
\jwst/NIRCam & F150W  & 27.052 & 1.120 & 1.096 & 0.979 & 30.1 \\
\hst/WFC3 IR & F160W  & 25.946 & 1.115 & 1.176 & - & 2.1  \\
\jwst/NIRCam & F200W  & 27.233 & 1.075 & 1.112 & 0.970 & 30.1 \\
\jwst/NIRCam & F277W  & 28.686 & 1.044 & 1.122 & 0.800 & 7.5  \\
\jwst/NIRCam & F356W  & 28.885 & 1.029 & 1.134 & 0.865 & 7.5  \\
\jwst/NIRCam & F444W  & 29.002 & 1.020 & 1.164 & 0.956 & 7.5  \\
\jwst/MIRI   & F770W  & 26.556 & 1.008 & 1.245 & - & 5.6  \\ 
\enddata
\tablecomments{For every \hst\ (ACS, WFC3) and \jwst\ (NIRCam, MIRI) filter at our disposal (columns 1 and 2), in this table we report its photometric zero-point ZPT$_{\rm AB}$ (column 3)\footnote{The {\it JWST} photometric zero-points reported in this table have been obtained from $-2.5\log_{10}(\rm PIXAR\_SR \cdot PHOTMJSR)-6.1$, and refer to the flux calibration used in the data reduction (\texttt{jwst\_0916.pmap}). To correct for the most updated version of the NIRCam flux calibration (\texttt{jwst\_0989.pmap}), the term $-2.5\log_{10}(f_{\rm ZPT})$ should be added to these values.}, the Galactic extinction correction factor $f_{\rm ext}$ (column 4), the aperture correction factor $f_{\rm aper}$ for a circular aperture of $1''$ diameter (column 5), the correction $f_{\rm ZPT}$ to account for the variation in the {\it JWST}/NIRCam photometric zero-points (\texttt{jwst\_0989.pmap}) with respect to the flux calibration used in the data reduction (\texttt{jwst\_0916.pmap}), and, finally, the total exposure time $t_{\rm exp}$ in ks (column 6). }
\label{tab:table_photometry}
\end{deluxetable}

\begin{figure}
    \centering
    \includegraphics[width=\columnwidth]{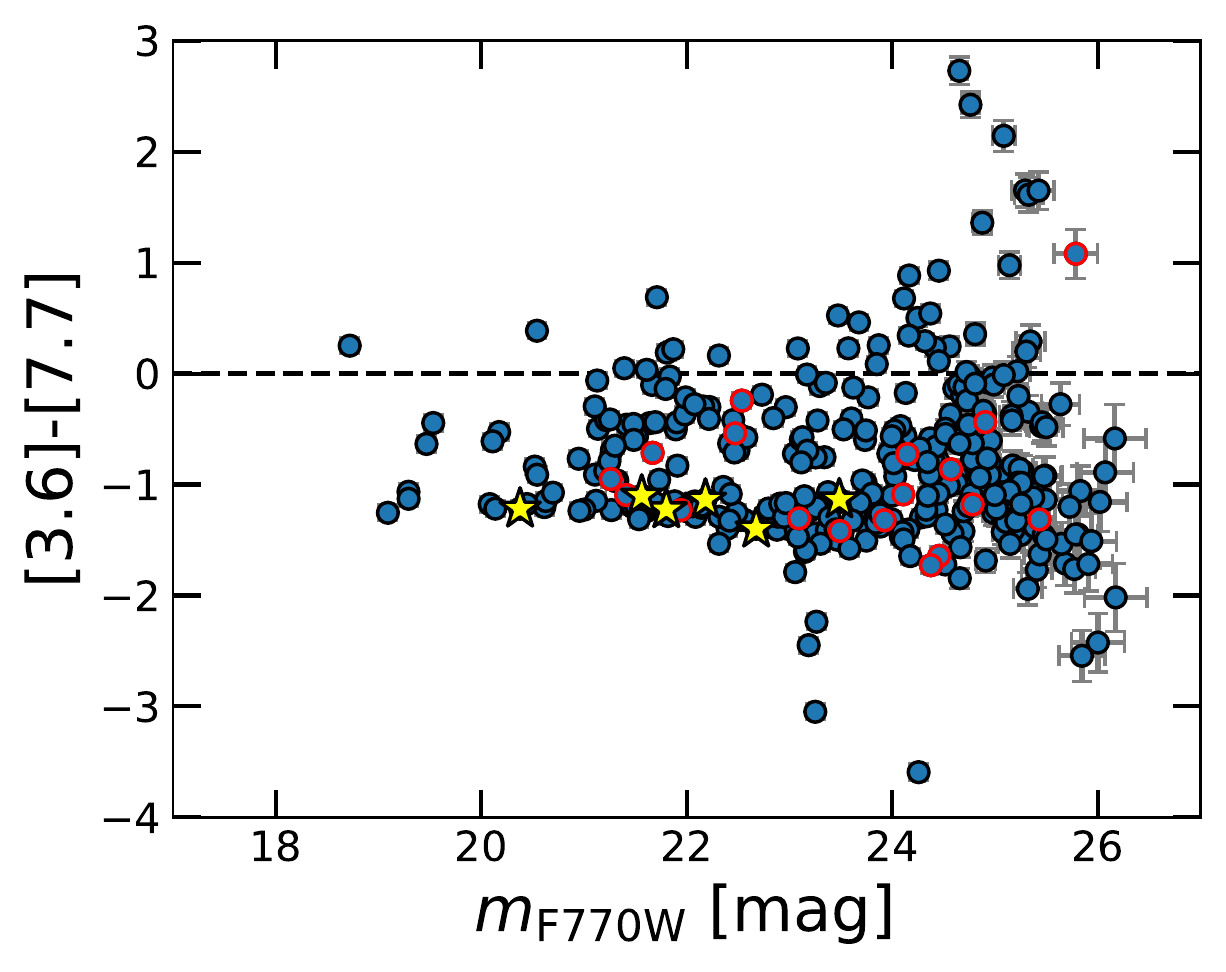}
    \caption{Colour magnitude diagram of our sample of 361 $7.7 \, \rm \mu m$-selected sources in SMACS 0723. The yellow stars are indicative of the 6 objects best-fitted by stellar templates. Finally, blue circles with red borders show the position of the objects discarded because of their high $\chi^2_{red}$.} 
    \label{fig:mag_cmd}
\end{figure}

Figure~\ref{fig:mag_cmd} shows the F356W -- F770W colour (hereafter $[3.6] - [7.7]$) of our sources as a function of their F770W magnitude. It is clear that most of the F770W sources are blue, with $[3.6]-[7.7] \sim -1$. A minority of sources ($\sim 9 \,\%$) are red instead ($[3.6]-[7.7]>0$), and these red sources appear mostly at $[7.7]>24$ mag.

\section{SED fitting and galaxy derived parameters} \label{sec:zphot}

We performed the spectral energy distribution (SED) fitting of our sources by means of the software \texttt{LePhare}\footnote{\url{https://www.cfht.hawaii.edu/~arnouts/LEPHARE/lephare.html}} \cite[][]{Arnouts+99, Ilbert+06}. For the SED fitting we considered only the photometry up to the NIRCam F444W filter, for a total of 13 \hst\ and NIRCam broad bands (see Figure \ref{tab:table_photometry}). We explicitly excluded the MIRI $7.7 \, \rm \mu m$ filter for the modelling because, a priori, we do not know the redshifts of our sources and, thus, we do not know whether the MIRI bands mainly trace stellar or dust emission.

We ran \texttt{LePhare} making use of the stellar population synthesis models from \cite{Bruzual+03} (hereafter BC03) based on Chabrier initial mass function \cite[IMF,][]{Chabrier+03}, and considering two-sets of stellar metallicities: solar ($Z_\odot = 0.02$) and sub-solar ($Z = 0.2Z_\odot$). We considered 
a range of star formation histories (SFHs): an instantaneous burst, i.e., a single stellar population (SSP) model, and a number of exponentially declining SFH ({\it $\tau$-model}), i.e. ${\rm SFR}(t) \propto e^{-t/\tau}$. 
For the $\tau$-models, we adopted values of $\tau$ (the so-called {\it e}-folding time) equal to 0.01, 0.1, 0.3, 1, 3, 5, 10, 15 Gyr.  We also complemented the BC03 stellar templates with the empirical QSO templates available in \texttt{LePhare} from \cite{Polletta+06}.
To take into account the effects of internal dust extinction, we convolved each synthetic spectrum with the attenuation law by \cite{Calzetti+00} and with the extrapolation proposed by \cite{Leitherer+02} at short wavelengths, leaving the colour excess E(B-V) as a free parameter with values ranging between $0.0 - 1.5$ in steps of 0.1. We ran \texttt{LePhare} in the redshift range $z = 0 - 18$ in a mode that takes into account the possible presence of nebular emission lines.

\subsection{Photometric Redshifts}

\begin{figure}
    \centering
    \includegraphics[width=\columnwidth]{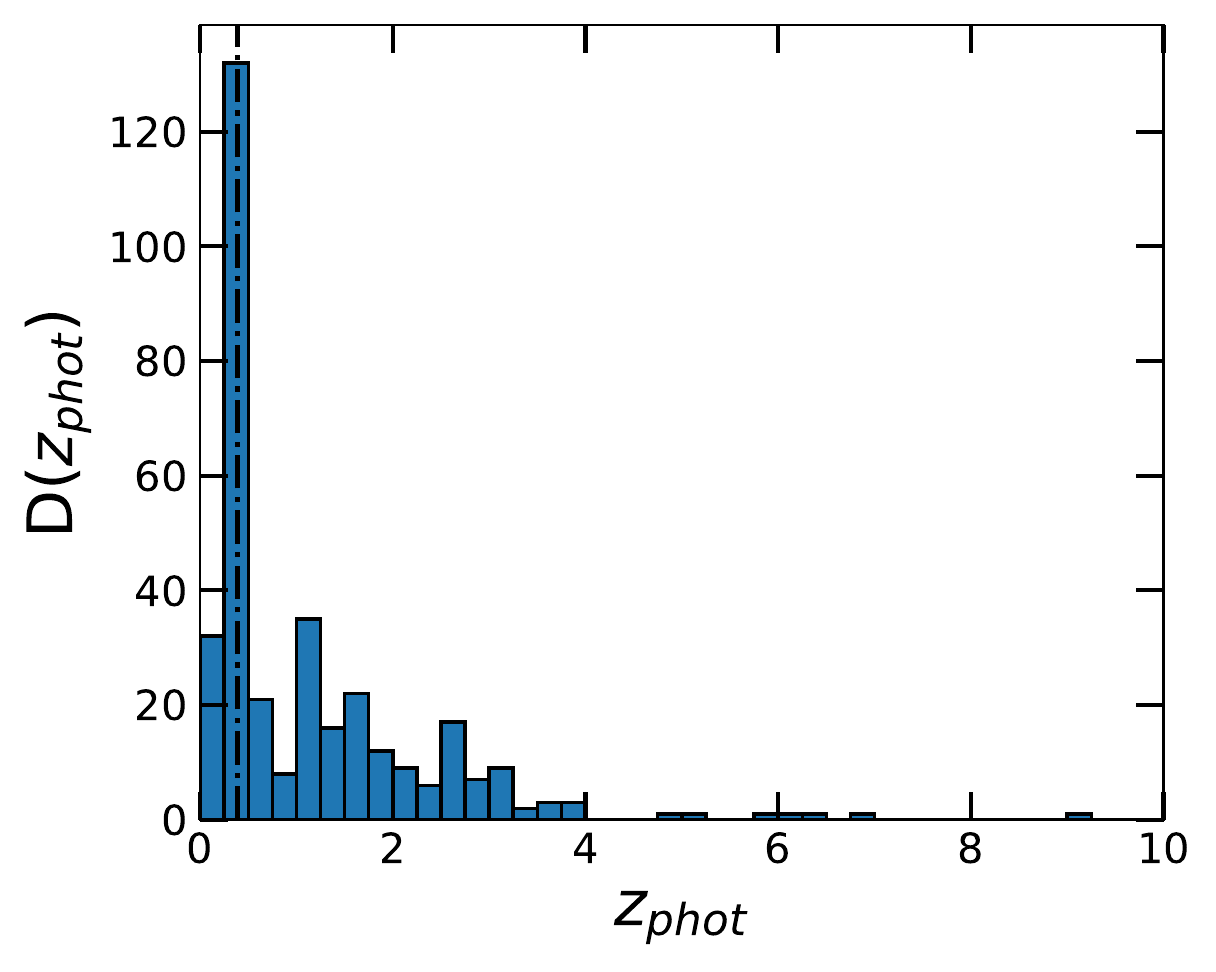}
    \caption{Photometric redshift distribution for our $7.7 \, \rm \mu m$-selected source sample in SMACS 0723. The vertical dash-dotted line indicates the redshift of the SMACS~0723 galaxy cluster ($z = 0.39$).}
    \label{fig:zphot_dist}
\end{figure}

From \texttt{LePhare}'s output\footnote{The complete photometric information as well as the physical properties of our final sample of 341 sources is presented in an electronic table available at \url{https://gitlab.astro.rug.nl/iani/smacs0723/}. In Table~\ref{tab:table_final} we present an excerpt of it, reporting the properties derived for the four sources at $z_{phot}>6$ we found in our sample (see Section~\ref{sec:z6ex}).} we obtained photometric redshift estimates, as well as derived physical parameters, for all our galaxies.
\texttt{LePhare} yielded a good SED fit for 96\% of all sources (median reduced chi-square $\chi_{red}^2 \simeq 1.09$). While, for the remaining 4\% (i.e. 14 objects), the code retrieved $\chi_{red}^2>20$ for all template solutions. We removed these 14 objects from all our subsequent analysis.
By comparing the $\chi^2_{red}$ of the different models, we find 6 objects with a clear preference for stellar templates ($\Delta \chi^2_{red}>4$). By looking at their F606W -- F814W vs F814W -- F356W colours \cite[e.g.][]{Caputi+11, Weaver+21}, these objects appear to fall along the stellar sequence. For this reason, we discarded these sources from our subsequent analysis.
The resulting redshift distribution of the remaining 341 sources is shown in Figure~\ref{fig:zphot_dist}.

We compared our photometric redshifts estimates with the secure spectroscopic redshifts\footnote{\url{https://wwwmpa.mpa-garching.mpg.de/~caminha/SMACS 0723_Caminha/model_JWST_v01/}} obtained by \cite{Caminha+22}, which were derived from observations with MUSE at the Very Large Telescope (VLT). 
Although the galaxies in our $7.7 \, \rm \mu m$ which are detected in MUSE are only $45$, we compared the photometric redshifts and spectroscopic values for the common sources.
Out of this matched sample, we identified 5 sources (11\%) as {\it catastrophic outliers}, i.e. $|z_{phot} - z_{spec}|/(1 + z_{spec}) > 0.15$. We derived a negligible bias for the $|z_{phot} - z_{spec}|/(1 + z_{spec}) \leq 0.15$ population (median($z_{phot}-z_{spec}$) = 0), and a tight $z_{phot} - z_{spec}$ correlation (standard deviation $\sigma = 0.04$). Overall, this confirmed the reliability of our photometric estimates.

As an additional check, we investigated the impact of possible secondary $z_{phot}$ solutions to our findings. According to \texttt{LePhare}, 14\% of our sample (49 objects) has a secondary redshift solution $z_{\rm SEC}$ such as $|z_{\rm BEST} - z_{\rm SEC}|>0.5$ and $|\chi^2_{red,{\rm BEST}}-\chi^2_{red,{\rm SEC}}|<4$. Nonetheless, if we chose to adopt the secondary $z_{phot}$ solution for these sources, we would not observe any substantial change in the overall redshift distribution of our sample nor for the trends on the derived physical quantities discussed in the following. This result suggests that our findings are robust against degeneracies in redshift space.

Figure~\ref{fig:zphot_dist} shows that the $7.7 \, \rm \mu m$ galaxy populations spans a very wide redshift range from $z=0$ through $z\sim 9.2$, with $57\%$ of the sources lying at $z<1$ and the remaining $43\%$ at higher redshifts. Sources above $z = 4$ constitute the 2\% of our sample.
The observed $7.7 \, \rm \mu m$ flux corresponds to rest-frame wavelengths which, at different redshifts, trace very different physical processes. At $z<1$, the observed  $7.7 \, \rm \mu m$ emission is dominated by the galaxy hot dust emission. At higher redshifts, the $7.7 \, \rm \mu m$ wavelength domain becomes increasingly dominated by stellar emission. Therefore, the breadth of the redshift distribution indicates the varied nature of the $7.7 \, \rm \mu m$ selected sources.

For the sources with an estimate of $z_{phot} > 1$ (145 objects in total), we decided to re-run \texttt{LePhare} adding the MIRI photometric measurement at 8$\mu$m. 
From this additional test, we confirmed the photometric redshift for 87\% of these sources (126 objects), while for the remaining 13\% (19 objects) we obtained a difference with the previous $z_{phot}$ estimate larger than 0.5. However, by comparing the $\chi^2_{red}$ of the two solutions, no objects were found to have a significantly better reduced chi-square, i.e. $\Delta\chi^2_{red} > 4$. Hence, we kept the best results obtained by fitting only the {\it HST} and {\it JWST}/NIRCam photometry.


\subsection{NIRCam/MIRI colours and galaxy internal extinctions}

\begin{figure}
    \centering
    \includegraphics[width=\columnwidth]{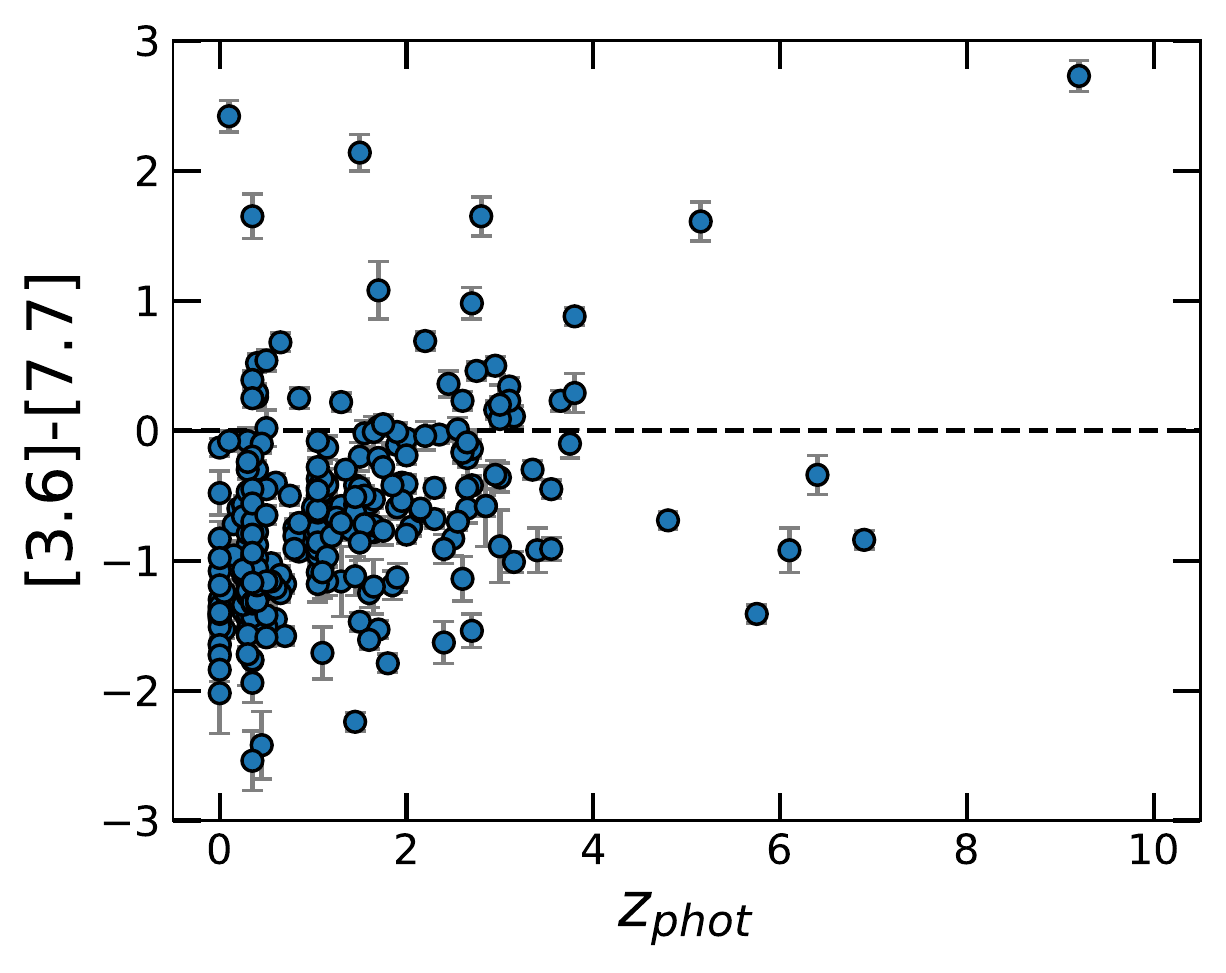}
    \caption{[3.6] - [7.7] versus photometric redshift $z_{phot}$ for our $7.7 \, \rm \mu m$-selected galaxy sample.}
    \label{fig:zphot_cmd}
\end{figure}

The varied nature of the $7.7 \, \rm \mu m$ source population is also illustrated in Figure~\ref{fig:zphot_cmd}. Even at similar redshifts, galaxies span a range of $[3.6]-[7.7]$ colours.

\begin{figure}
    \centering
    \includegraphics[width=.96\columnwidth]{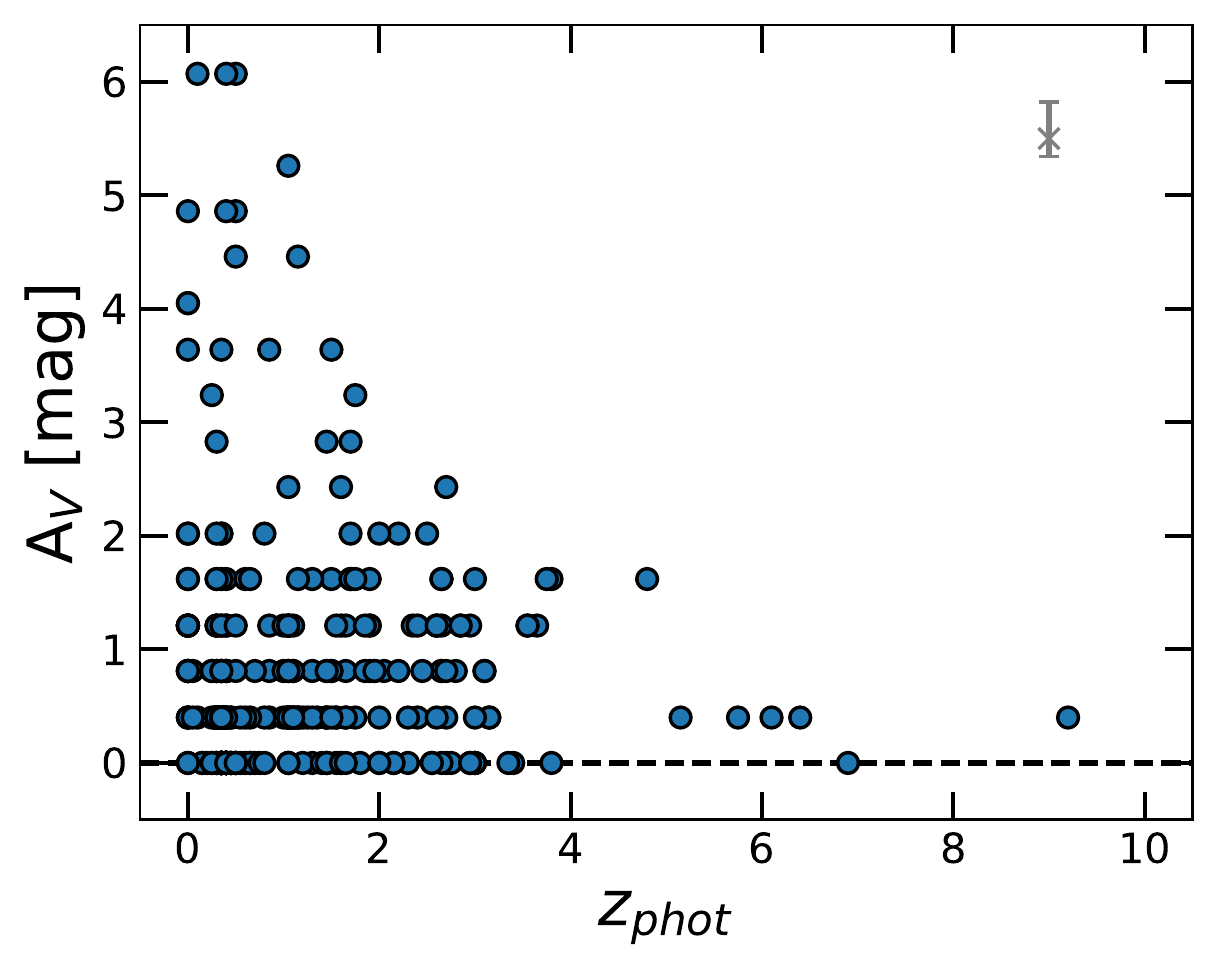}
    \caption{Dust extinction A$_V$ as a function of photometric redshift $z_{phot}$. In the top right corner we report in gray the typical (median) value of the error associated to the A$_V$ estimate.}
    \label{fig:av_diagram}
\end{figure}

The best-fit SEDs of the $7.7 \, \rm \mu m$ sources are characterised by a wide range of $A_V$ values (Figure~\ref{fig:av_diagram}). Only a minority ($<10\%$) require high dust extinctions ($A_V>2$~mag) for the SED fitting and are virtually all sources at $z<3$ in our sample. Interestingly, the sources with red colour $[3.6]-[7.7]>0$ do not preferentially correspond to cases with high $A_V$ values.

\cite{Stern+05} proposed to isolate potential AGN using IRAC [3.6]-[4.5] versus [5.8]-[8.0] colours. Unfortunately we do not have MIRI 5.6$\, \rm \mu m$ photometry for SMACS 0723, but we can analyse [3.6]-[4.4] colours instead. We find that we only have a few sources with [3.6]-[4.4]$>0$ at $z>2$, so AGN with an IR power-law do not seem to be a significant fraction of our high-$z$ sample (the surface density of these sources is likely too low to see them in significant numbers in the small area covered by SMACS 0723).

\subsection{Stellar masses of $z>1$ galaxies}
\label{subsec:stellar_mass}

\begin{figure}
    \centering
    \includegraphics[width=\columnwidth]{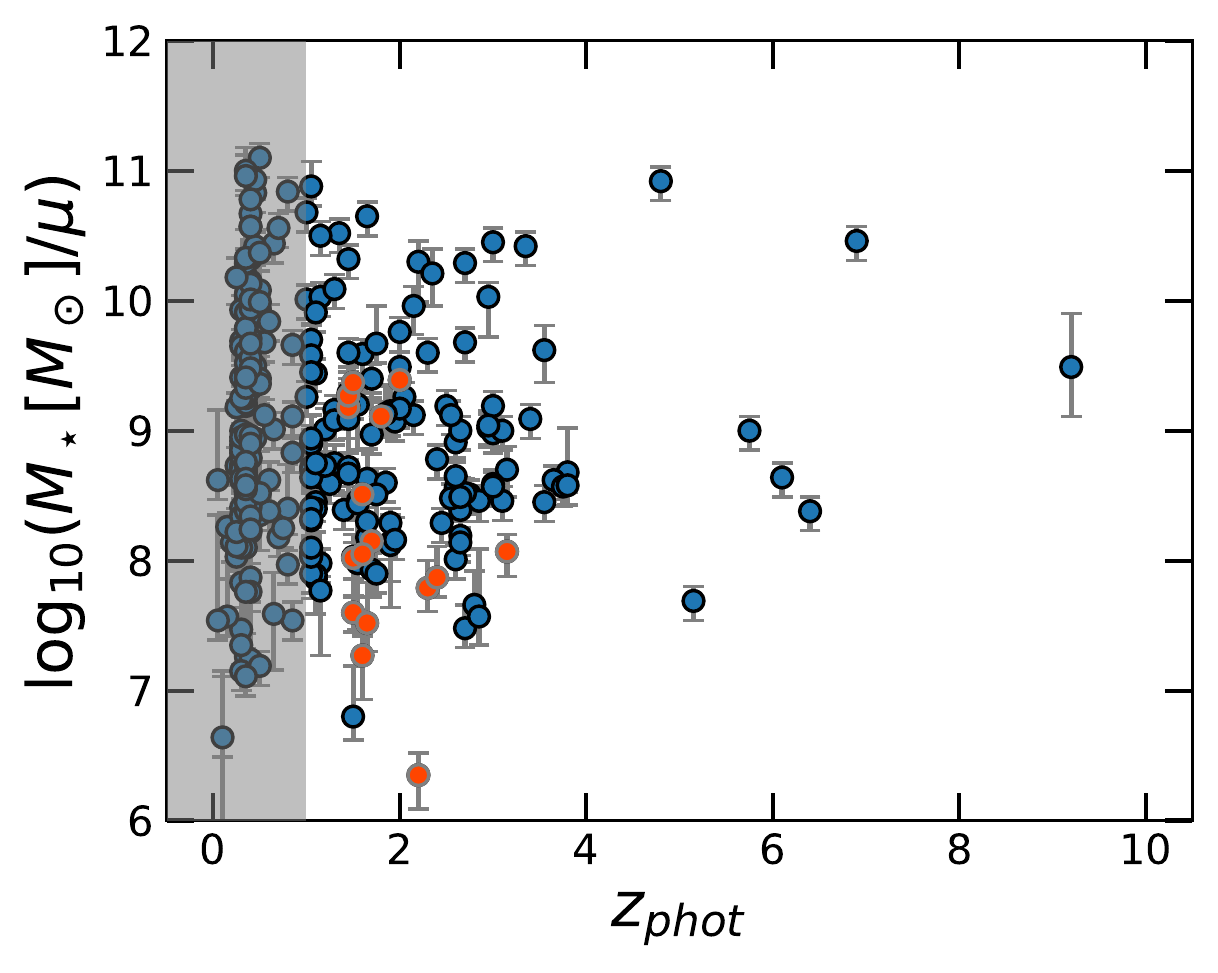}
    \caption{Diagram of the stellar mass (corrected for lensing magnification) of our sample of sources as a function of the photometric redshift $z_{phot}$. The orange circles identify the objects with a magnification factor $\mu$ above 10, while the shaded area highlights the region of the diagram where $z_{phot}<1$.}
    \label{fig:mass-zphot}
\end{figure}

For completeness, we also analyse the derived stellar masses for our galaxies. As SMACS 0723 is a lensing cluster, the stellar masses of background sources need to be corrected by lensing magnification. We used the lensing model by \cite{Caminha+22} (see Section~\ref{subsec:lensing}) to correct these stellar masses for magnification. 
The lensing-corrected stellar mass versus $z_{phot}$ for all our galaxies at $z>1$ are shown in Figure~\ref{fig:mass-zphot}.
The median magnification factor for our sample is $\mu = 2.56$. About 11\% of sources above $z>1$ (i.e. 5\% of the overall sample) have a magnification factor above 10 and they are all objects with $z_{phot} < 4$ (see Figure~\ref{fig:mass-zphot}).

The 7.7~$\rm \mu m$ source sample contains galaxies with a wide range of stellar masses $10^7$-$10^{11} \, \rm M_\odot$ at $z>1$, indicating that a selection in this waveband does not necessarily correspond to a selection in stellar mass even at $z>1$.  Nevertheless, it should be noted that the sparse population of the stellar-mass plot at $z>4$ is to some extent due the fact that these galaxies are increasingly rarer towards higher redshifts and the area sampled by SMACS~0723 is small. So not all stellar-mass galaxies at high $z$ might be represented in this field due to sample variance, but not because of a consequence of the 7.7~$\rm \mu m$ selection.

This is also true for the highest redshift sources at $z > 6$ in our sample. A few of them have stellar masses
$10^8-10^{9}\, \rm M_\odot$, while one has $\sim 2\times10^{10}\, \rm M_\odot$, i.e., it is quite massive for such high redshifts (see \S\ref{sec:z6ex}).
In fact, even though galaxies with stellar masses $M_\star \sim 10^{10}~\rm M_\odot$ at $z \sim 7$ have been previously found with {\it Spitzer}, they were only detected in IRAC channels 1 and 2 \cite[e.g][]{Roberts-Borsani+16, Mierlo+22}, not at longer wavelengths.

\subsection{$7.7 \, \rm \mu m$ sources at $z>6$}
\label{sec:z6ex}

\begin{figure*}
    \centering
    \includegraphics[height=.3\textheight]{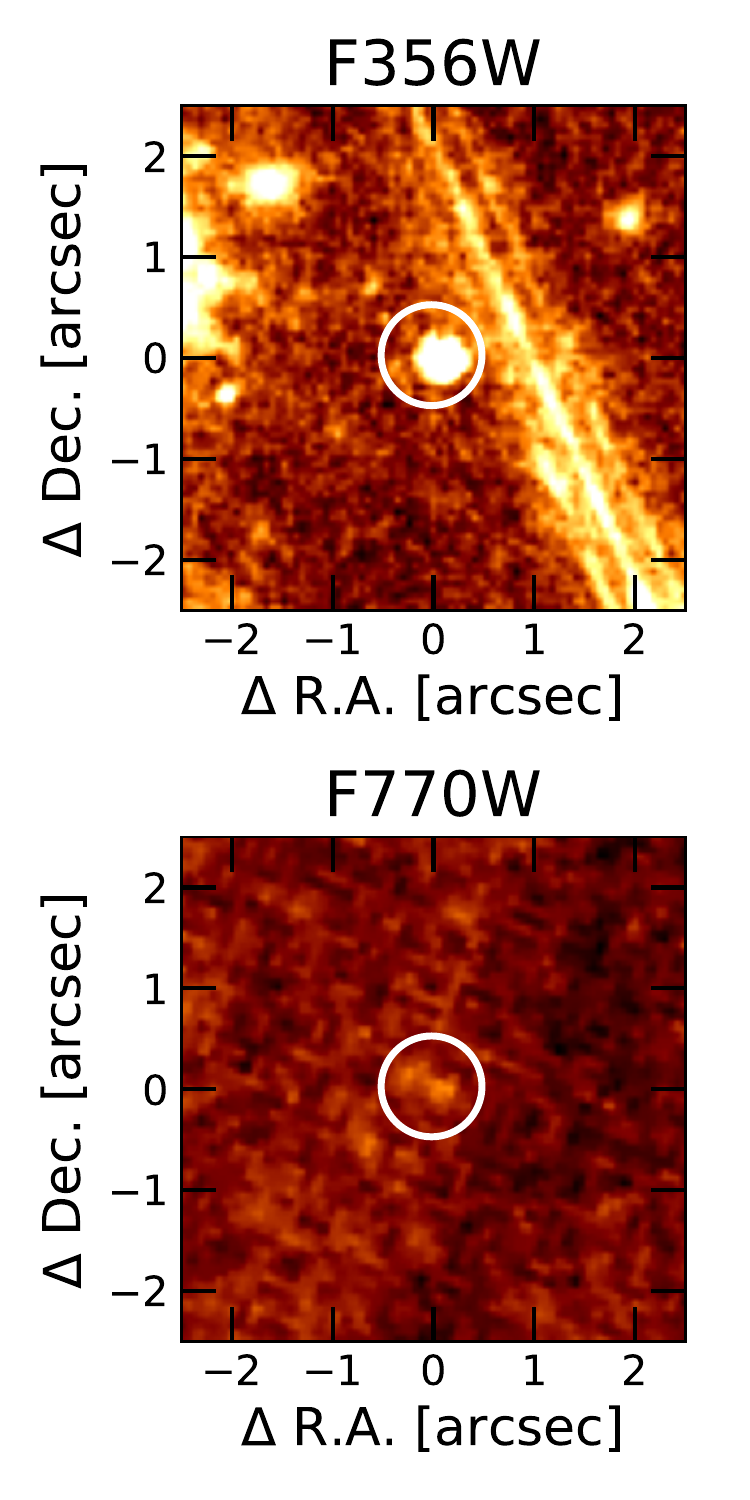}
    \includegraphics[height=.32\textheight]{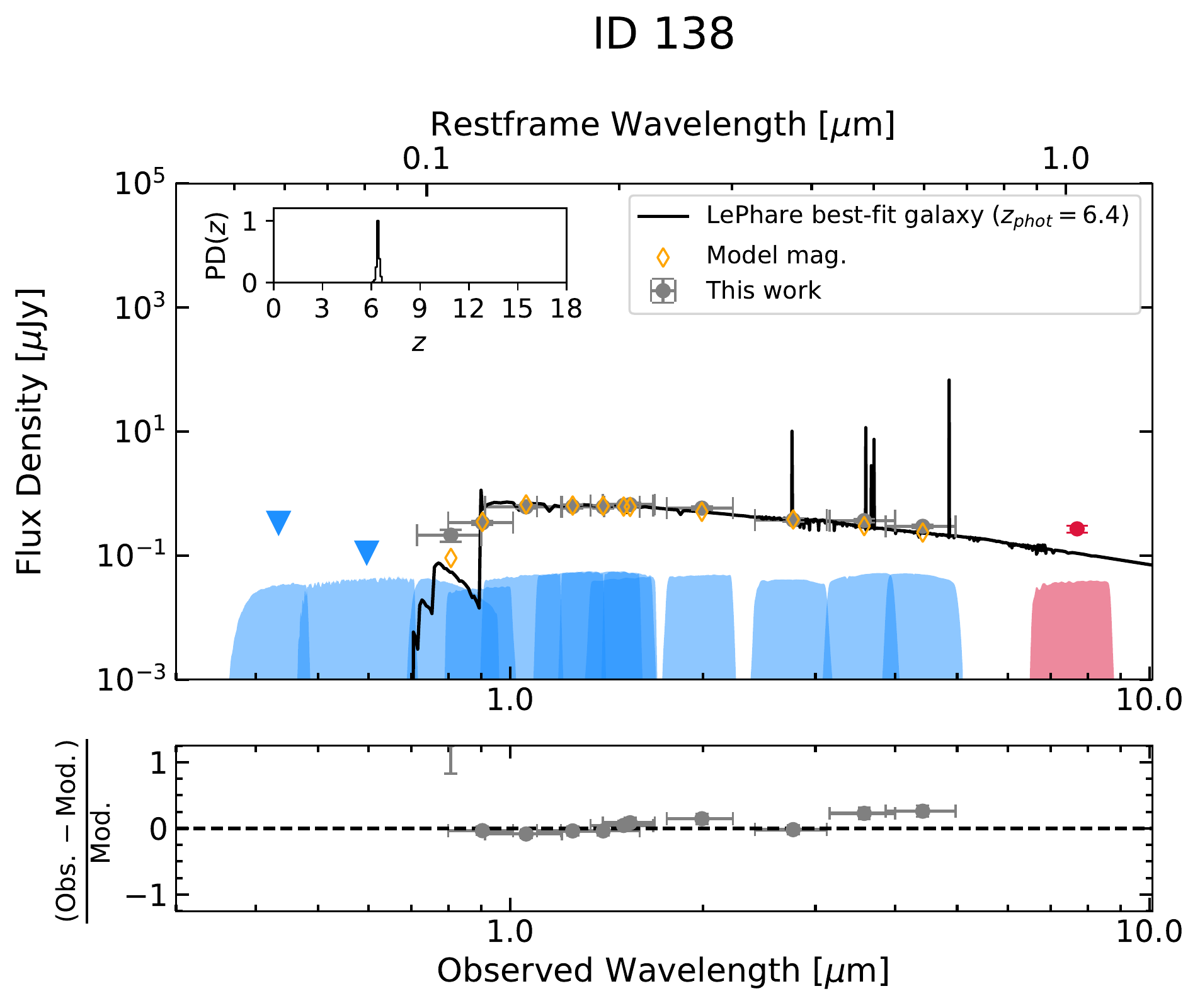}\\
    \medskip
    \includegraphics[height=.3\textheight]{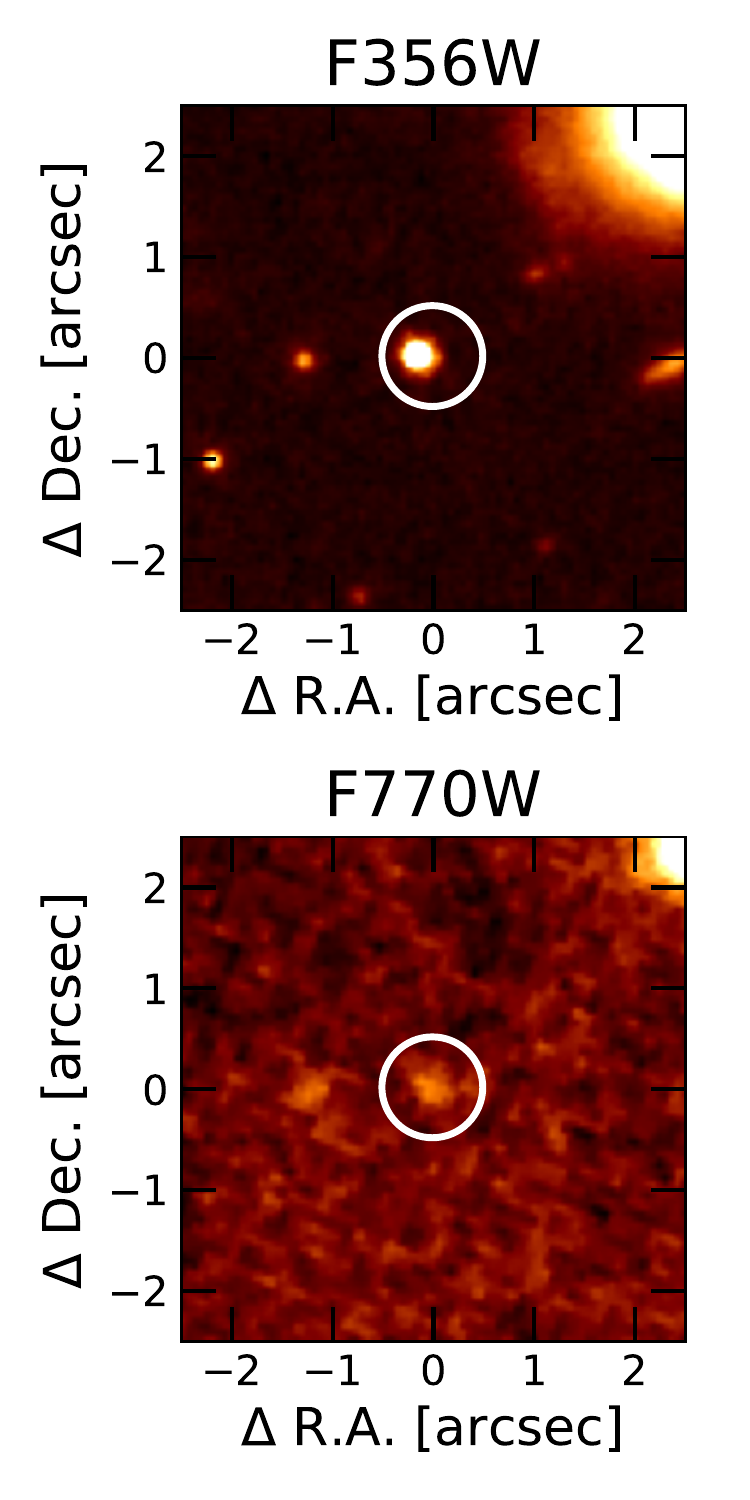}
    \includegraphics[height=.32\textheight]{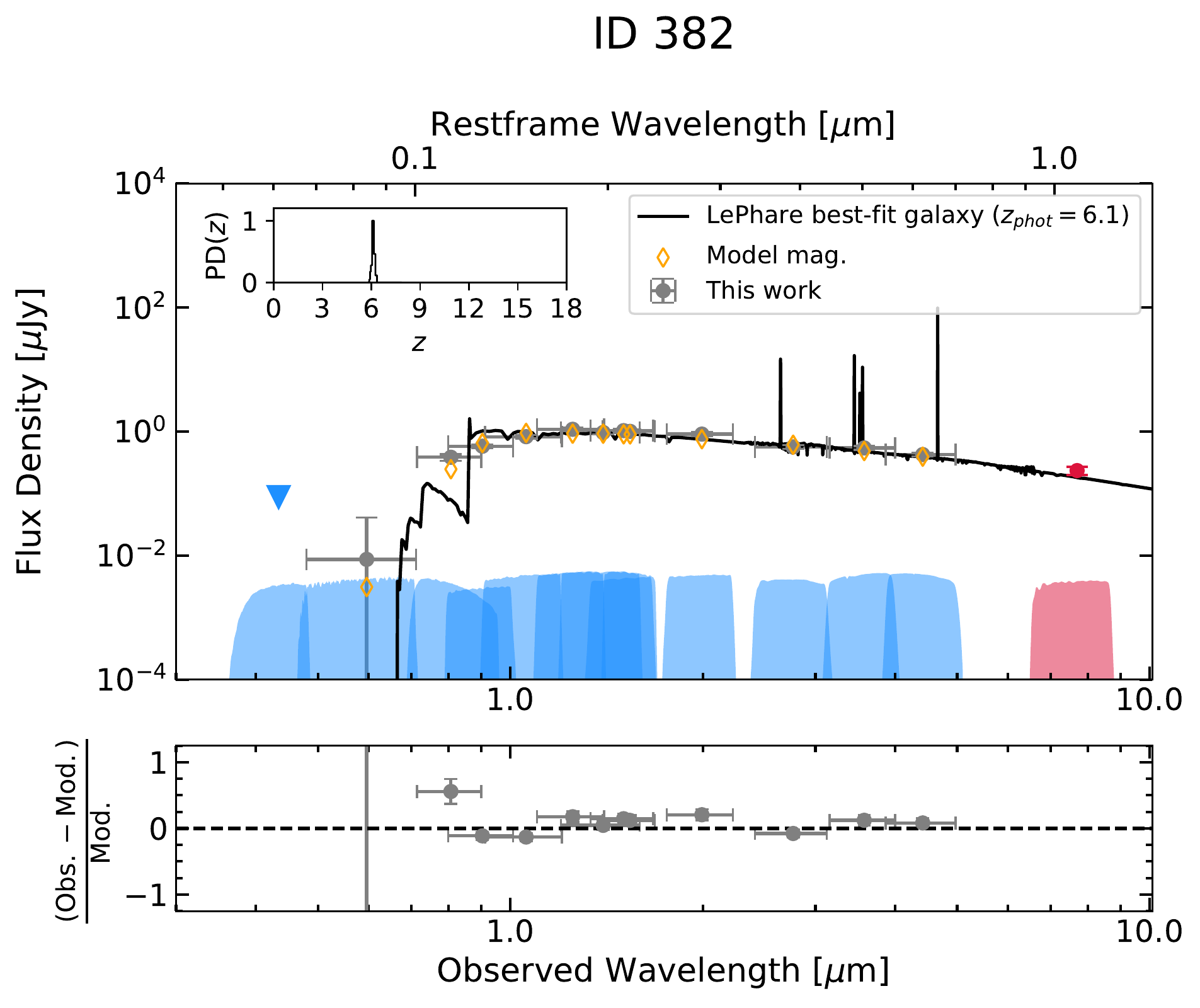}\\
    \medskip
    \includegraphics[height=.3\textheight]{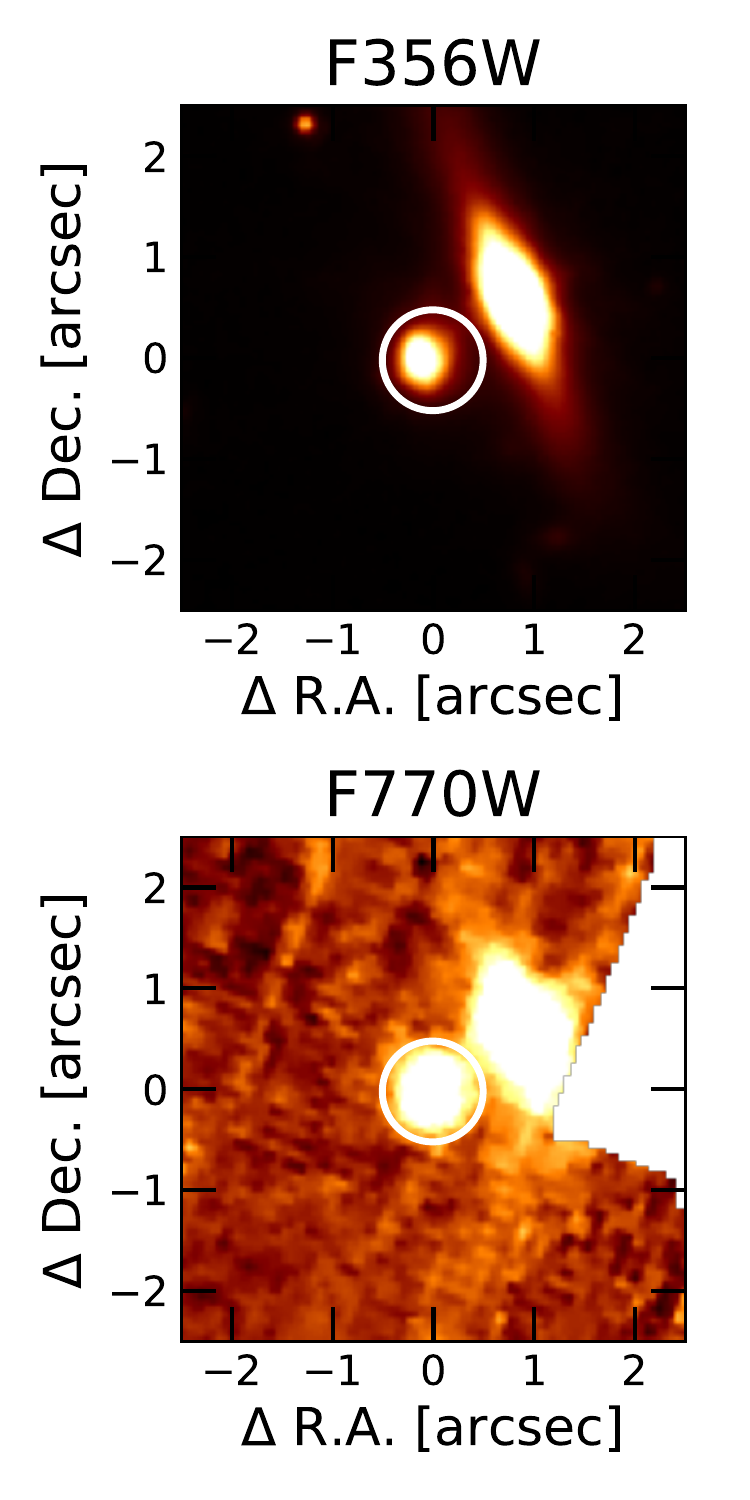}
    \includegraphics[height=.32\textheight]{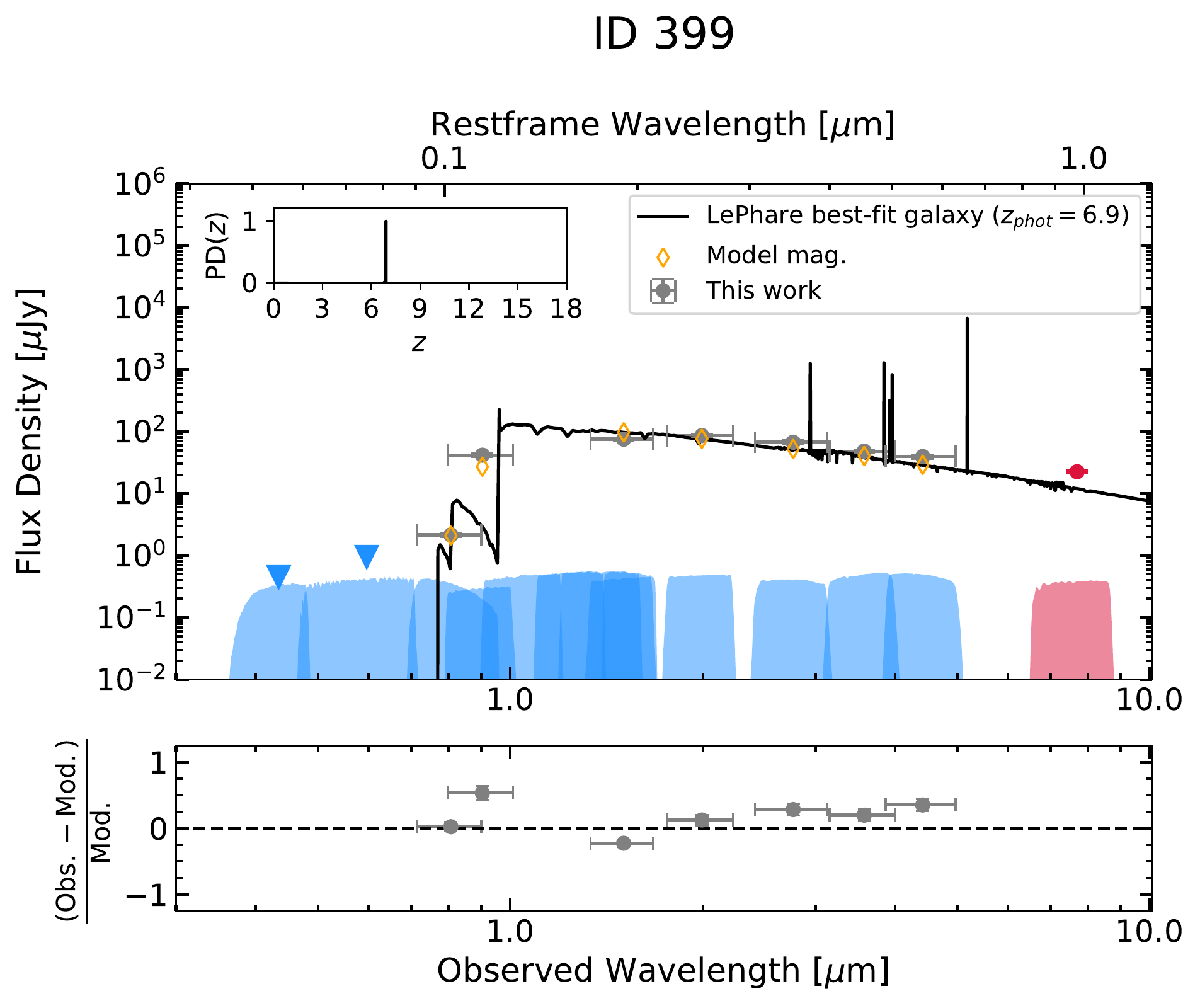}\\
    \caption{\texttt{LePhare} best-fit SED of galaxies ID 138 (top), ID 382 (centre) and ID 399 (bottom), both at $z_{phot}>6$. The MIRI $7.7 \, \mu m$ point (in red) has not been used for the SED fitting, but added a posteriori. The inset on the top-left corner shows the probability distribution for the estimate of the photometric redshifts $z_{phot}$ (PDZ). For the three galaxies, we show cutout images ($5''\times 5''$) in the NIRCam F356W and MIRI F770W bands. }
    \label{fig:seds}
\end{figure*}

\begin{figure*}
    \centering
    \includegraphics[width=.24\textwidth]{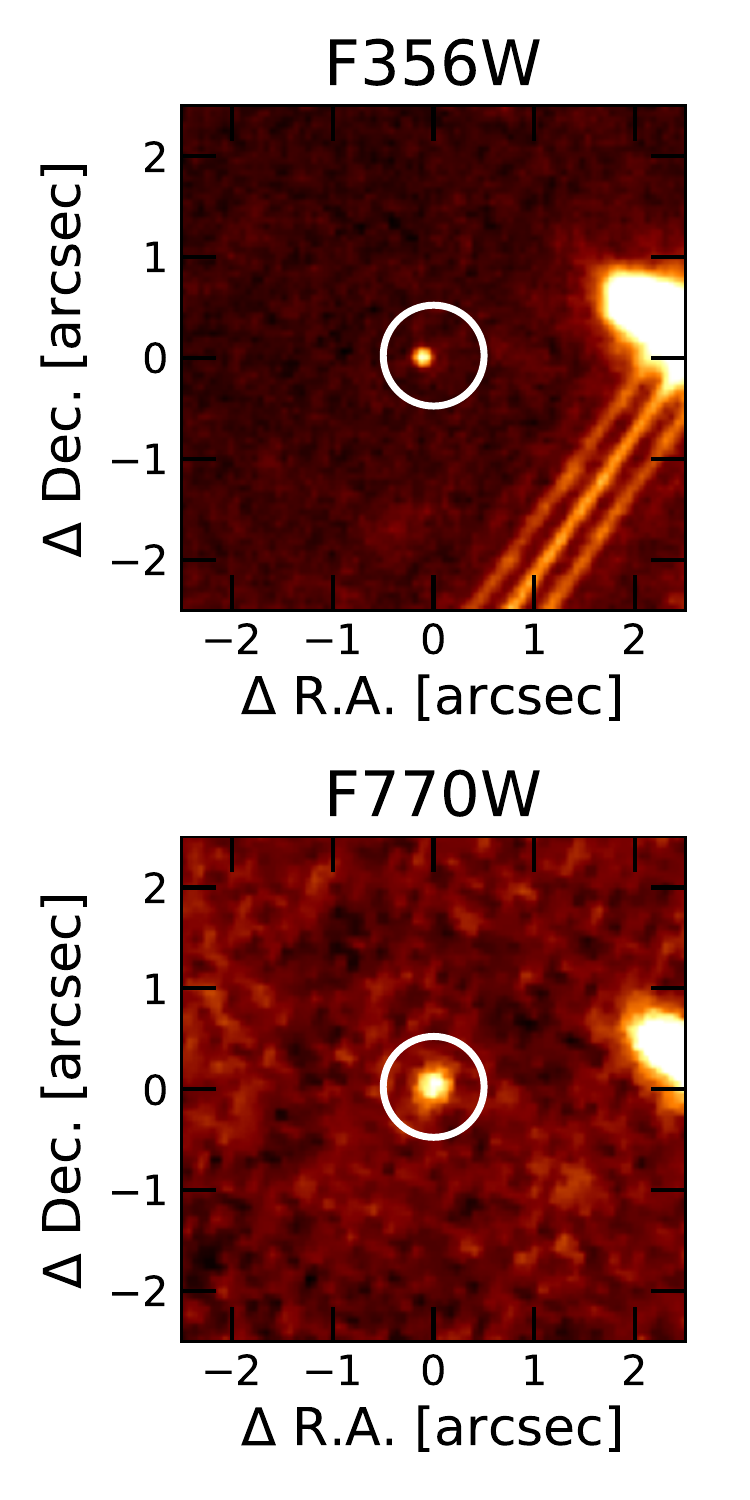}
    \includegraphics[width=.64\textwidth]{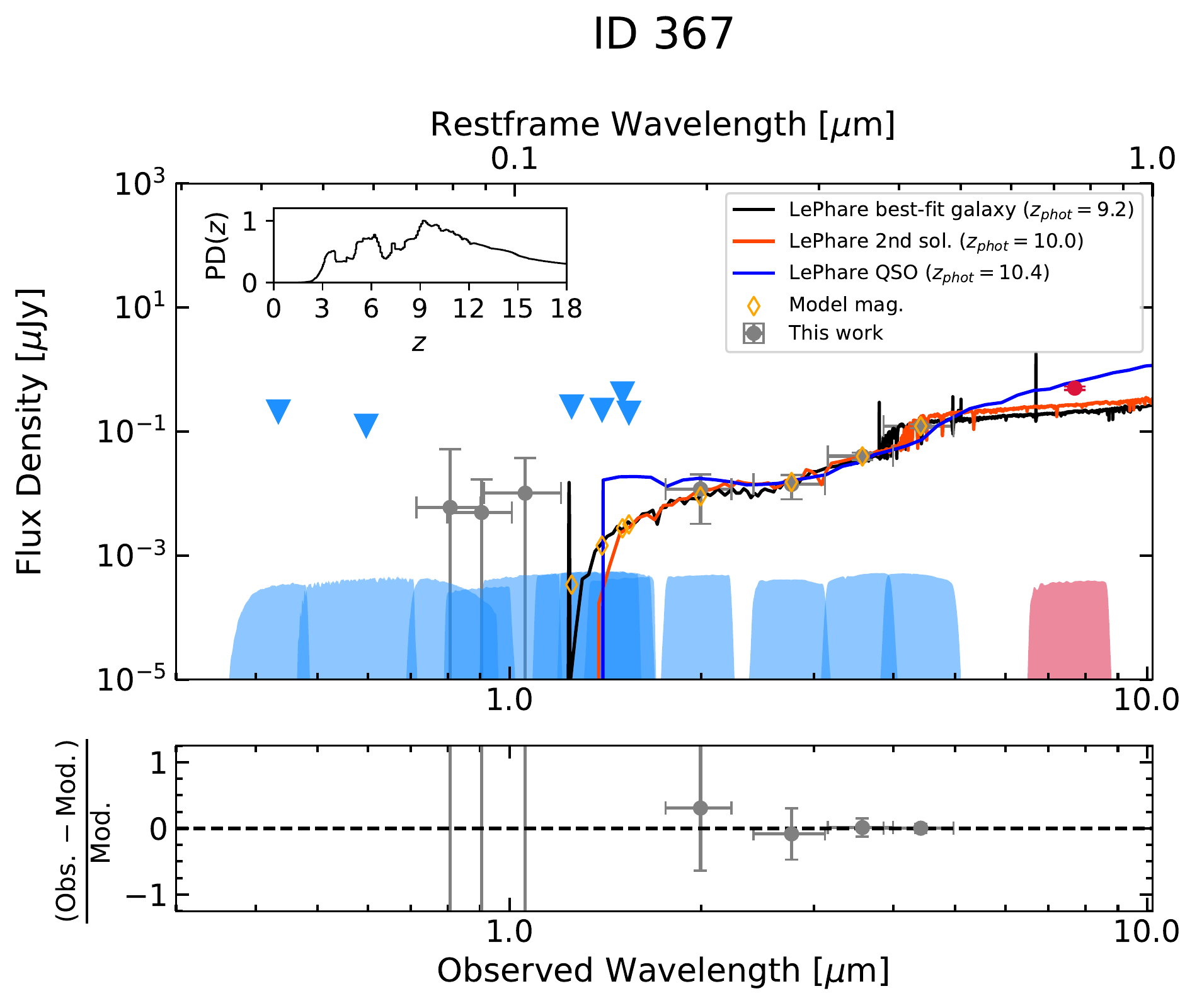}\\
    \caption{\texttt{LePhare} best-fit SED of galaxy ID 367 at $z_{phot}=9.2$ (in black). The orange and blue spectra show \texttt{LePhare} best-fit secondary galaxy solution ($z_{phot}=10$) and QSO template ($z_{phot}=10.4$), respectively. The MIRI $7.7 \, \mu m$ point has not been used for the SED fitting, but added a posteriori. The inset on the top-left corner shows the PDZ of this source. On the left-hand side panels, we show cutout images ($5''\times 5''$) of ID 367 in the NIRCam F356W and MIRI F770W bands.}
    \label{fig:id367}
\end{figure*}

Figure~\ref{fig:seds} and Figure~\ref{fig:id367} show postage stamps and the best-fit SEDs of all our galaxies at $z_{phot}>6$. The postage stamps show their clear detection at both 7.7 and $3.6 \, \rm \mu m$. In both cases, the resulting SED fitting is of very good quality. For ID 382, even the $7.7 \, \rm \mu m$ photometric point, which was not used for the SED fitting, is in excellent agreement with the extrapolation of the best-fit template at observed wavelengths $>5 \, \rm \mu m$.

Source ID 138 at $z_{phot} = 6.4$ has a best-fit dust extinction $A_V = 0.4$ mag and a delensed stellar mass of $2.4 \times 10^{8} \, \rm M_\odot$ (magnification factor $\mu = 4.02 \pm 0.46$). Similarly, ID 382 at $z_{phot} = 6.1$ ($\mu = 2.86 \pm 0.13$) has $A_V = 0.4$ mag and delensed stellar mass of $4.3 \times 10^{8} \, \rm M_\odot$ . ID 399 at $z_{phot} = 6.9$ ($\mu = 2.32 \pm 0.05$), instead, has a visual extinction $A_V = 0$ mag and a delensed stellar mass $\sim 2\times 10^{10}\, \rm M_\odot$.
Interestingly, all these sources are not affected by strong magnification ($\mu \sim 2-4$), thus suggesting that future deeper MIRI observations at $7.7. \mu$m will be able to detect such objects even in non-lensed fields.
In Table~\ref{tab:table_final}, we report all the photometric information we derived for these sources as well as their inferred physical properties. 

The last object with $z_{phot}>6$ is ID 367. 
This is the source with the highest photometric redshift estimate out of all our sample, i.e. $z_{phot} = 9.2$. 
As shown in Figure~\ref{fig:id367}, this source has a complex redshift probability distribution (PDZ), with a secondary photometric solution at $z_{phot}=10$ and a solution from the QSO models with $z_{phot}=10.4$. 
Despite the best-fit models prefer high-$z$ solutions, low-$z$ solutions cannot be completely ruled out as pointed out by the breadth of the PDZ which extends down to $z \sim 3$.  
The SED of this object shows a clear and steep rise above 2$\mu$m and, interestingly, the QSO template seems to nicely predict the observed 8 $\mu$m flux of this source. 
Nonetheless, based on the $\chi^2_{red}$, \texttt{LePhare} does not present a clear preference among the different best-fit models ($|\Delta \chi^2_{red}| \lesssim 0.2$ in all cases). 
Besides, by re-running the SED fitting considering the MIRI F770W filter, the photometric redshift solutions do not appear to vary significantly ($\Delta z_{phot}\leq 0.5$) and, at the same time, the $\chi^2_{red}$ does not improve ($|\Delta \chi^2_{red}|\lesssim 0.3$). 
For this object, we estimate an $A_V = 0.4$ mag and a stellar mass of $3 \times 10^9 \, \rm M_\odot$. 
ID 367 was also reported by \cite{Rodighiero+22} (dubbed in their catalogue as \texttt{PENNAR}) with a primary photometric solution $z_{phot} = 12.1$, a visual extinction $A_V \simeq 2.5$ mag and a stellar mass $\sim 4\times 10^9 \, \rm M_\odot$. 
We ascribe the differences in the results as due to the diverse technique of SED fitting performed, the differences in the photometric NIRCam zero-points applied, as well as to the fact that we do not implement the MIRI longer-wavelength filters ($>8 \mu$m) in our SED fit.

\begin{splitdeluxetable*}{lcccccccccBcccccccBccccccc}
\rotate
\tablenum{2}
\tablecaption{Main properties of our sample of $7.7\mu$m-detected sources at $z_{phot}>6$ (excerpt of the electronic table reporting the properties of the 341 sources of our final sample).}
\tablewidth{0pt}
\tablehead{
\colhead{ID} & \colhead{R.A.} & \colhead{Dec.} & \colhead{$f_{\rm F435W}$} & \colhead{$f_{\rm F606W}$} & 
\colhead{$f_{\rm F814W}$} & \colhead{$f_{\rm F090W}$} & \colhead{$f_{\rm F105W}$} & 
\colhead{$f_{\rm F125W}$} & \colhead{$f_{\rm F140W}$} & \colhead{$f_{\rm F150W}$} & 
\colhead{$f_{\rm F160W}$} & \colhead{$f_{\rm F200W}$} & \colhead{$f_{\rm F277W}$} & 
\colhead{$f_{\rm F356W}$} & \colhead{$f_{\rm F444W}$} & \colhead{$f_{\rm F770W}$} & 
\colhead{${\rm [3.6] - [7.7]}$} & \colhead{$z_{\rm BEST}$} & \colhead{$z_{\rm SEC}$} & \colhead{A$_V$} & \colhead{$\chi^2_{\rm red, BEST}$} & $\mu$ & \colhead{$\log_{10}({\rm M}_{\star, \rm delens})$}\\
 & [deg] & [deg] & [nJy] & [nJy] & [nJy] & [nJy] & [nJy] & [nJy] & [nJy] & [nJy] & [nJy] & [nJy] & [nJy] & [nJy] & [nJy] & [nJy] & & & & [mag] & & & [M$_\odot$]}
\startdata
138 & 110.862082 & -73.462243 & $339.99^\dagger$ & $112.44 ^\dagger $ & $215.02 \pm 31.01$ & $340.10 \pm 15.66 $ & $ 614.75 \pm 28.31$ & $620.47 \pm 30.59 $ & $ 613.09 \pm 28.23 $ & $ 648.40 \pm 29.86 $ & $ 666.24 \pm 30.68 $ & $ 586.16 \pm 27.00 $ & $ 377.19 \pm 17.37 $ & $ 369.25 \pm 17.01 $ & $ 298.45 \pm 13.75 $ & $ 269.09 \pm 34.33 $ & $ -0.34 \pm 0.15 $ & 6.40 & - & $0.40_{-0.12}^{+0.20}$ & 1.96 & $4.02 \pm 0.46$ & $8.38_{-0.15}^{+0.11}$ \\ 
367 & 110.819567 & -73.444912 & $211.97^\dagger$ & $125.44^\dagger $ & $6.01 \pm 31.03$ & $4.98 \pm 7.99 $ & $ 10.21 \pm 18.33$ & $255.49^\dagger $ & $ 226.05^\dagger $ & $ 414.71^\dagger $ & $ 203.24^\dagger $ & $ 11.91 \pm 5.75 $ & $ 14.17 \pm 3.98 $ & $ 40.42 \pm 3.66 $ & $ 123.55 \pm 5.69 $ & $ 499.57 \pm 34.36 $ & $+ 2.73 \pm 0.12 $ & 9.2 & 10.0 & $0.40_{-0.40}^{+0.81}$ & 1.07 & $3.60 \pm 0.20 $ & $9.49_{-0.38}^{+0.41}$ \\ 
382 & 110.842626 & -73.444114 & $88.35^\dagger$ & $8.74 \pm 21.55$ & $387.53 \pm 31.05$ & $583.23 \pm 26.86 $ & $ 832.42 \pm 38.33$ & $1095.27 \pm 50.44 $ & $ 973.58 \pm 44.83 $ & $ 1041.02 \pm 47.94 $ & $ 1013.60 \pm 46.68 $ & $ 915.56 \pm 42.17 $ & $ 568.88 \pm 26.20 $ & $ 548.16 \pm 25.24 $ & $ 428.27 \pm 19.72 $ & $ 234.05 \pm 34.36 $ & $ -0.92 \pm 0.17 $ & 6.1 & - & $0.40_{-0.12}^{+0.16}$ & 3.66 & $2.86 \pm 0.13$ & $8.64_{-0.15}^{+0.11}$ \\ 
399 & 110.776827 & -73.442334 & $452.88^\dagger$ & $959.86^\dagger $ & $2167.61 \pm 99.82$ & $41870.86 \pm 1928.23$ & $ - $ & $ - $ & $ - $ & $ 75889.15 \pm 3494.82 $ & $ - $ & $ 86764.42 \pm 3995.65 $ & $ 67527.70 \pm 3109.77 $ & $ 48620.38 \pm 2239.05 $ & $ 39783.05 \pm 1832.08 $ & $ 22461.64 \pm 1034.40 $ & $ -0.84 \pm 0.07 $ & 6.9 & - & $0.0$ & 2.13 & $2.32 \pm 0.05$ & $10.46_{-0.14}^{+0.11}$ \\
\enddata
\tablecomments{Table 2 is published in its entirety in the machine-readable format. A portion is shown here for guidance regarding its form and content. All the physical parameters presented and discussed in this paper for our final sample of 341 $7.7\mu$m-detected sources in SMACS 0723 are available at \url{https://gitlab.astro.rug.nl/iani/smacs0723/}. In this table, $^\dagger$ is indicative of photometric measurements for which we have only an upper-limit.}
\label{tab:table_final}
\end{splitdeluxetable*}

\section{Summary and Conclusions} \label{sec:conclusion}

In this paper we have presented the first physical-parameter characterisation of the 7.7~$\rm \mu m$-selected galaxy population based on \jwst\ MIRI and NIRCam data. We studied the lensing cluster  SMACS 0723, which was one of the first extragalactic fields observed with \jwst. We complemented the \jwst\ data with ancillary \hst\ images, allowing us to perform the galaxy SED fitting on a total of 13 broad bands for our total population of 341 $7.7 \, \rm \mu m$-selected galaxies with $S/N>3$ in both F770W and F356W.  

We obtained photometric redshifts and other derived physical parameters for all our sources. We showed that at the depth of the current MIRI survey, the $7.7 \, \rm \mu m$ selection results in a heterogenous galaxy population: $57\%$ of the selected galaxies are at $z<1$ for which the $7.7 \, \rm \mu m$ light should be mostly produced by hot dust emission. The other $43\%$ corresponds to galaxies at $z>1$, a redshift regime where the $7.7 \, \rm \mu m$ light is increasingly dominated by stellar emission. Our high-$z$ sample has a small tail (2\%) of objects at $z>4$, which reaches $z=9.2$, indicating that already at these depths the MIRI $7.7 \, \rm \mu m$  images contain galaxies within the epoch of reionisation.

All the objects in our final sample are best fit with galaxy templates. In addition, one of our galaxies (ID 146) at $z_{phot}=2.7$ is an ALMA millimetre source observed in the ALCS program \cite[P.I. K. Kohno; see ][]{Sun+22}.

The vast majority of the 7.7~$\rm \mu m$ galaxies have blue $[3.6]-[7.7]$ colours, independently of redshift.  Less than 10\% of them have red colours $[3.6]-[7.7]>0$ and this red colour is not necessarily associated with high $A_V$ values. Although the 7.7~$\rm \mu m$ galaxies span a wide range of best-fit $A_V$ values, only a few of them at $z<2$ seem to need the maximum $A_V\approx 6$ mag permitted in our SED-fitting runs.

The $z>1$ galaxies in our sample span stellar masses between $10^7$ and $10^{10} \, \rm M_\odot$.  The access to low stellar masses is facilitated thanks to the gravitational lensing effect, as it has been shown by previous works \cite[e.g.,][]{Rinaldi+22}. We do not find any very massive $>10^{11} \, \rm M_\odot$ galaxy, but this is not surprising, given the small volume sampled by the SMACS~0723 field.
Concurrently, we found that the extension of our sample to high-$z$ sources ($z>6$) is not driven by lensing magnification (average magnification factors $\mu\simeq 2-4$). This suggests that deeper MIRI $7.7 \mu$m observations will be able to detect such high-$z$ objects even in non-lensing fields.

Overall, this work demonstrates the enormous potential of MIRI to open up the study of the high-redshift Universe at mid-IR wavelengths, particularly at $7.7 \, \rm \mu m$ where MIRI still has relatively high sensitivity. Very soon, deeper  MIRI $7.7 \, \rm \mu m$ observations will be available, allowing us to investigate a larger number of high-$z$ galaxies, all the way into the epoch of reionisation. 

\begin{acknowledgements}
The authors thank the anonymous referee for the constructive suggestions that helped improving the manuscript.
EI and KIC acknowledge funding from the Netherlands Research School for Astronomy (NOVA).
KIC and VK acknowledge funding from the Dutch Research Council (NWO) through the award of the Vici Grant VI.C.212.036. The Cosmic Dawn Center is funded by the Danish National Research Foundation under grant No. 140. We thank Kotaro Kohno for discussion on the ALMA ALCS sources in SMACS~0723.
\end{acknowledgements}


\bibliography{biblio}{}
\bibliographystyle{aasjournal}



\end{document}